\documentclass[preprint,12pt,authoryear]{elsarticle}

\usepackage{amssymb, amsmath}
\usepackage{mathtools}

\usepackage{bm}
\usepackage{xcolor}
\usepackage{url}

\usepackage{tikz}
\usetikzlibrary{positioning}

\newcommand{\CV}{\mathcal{V}}

\journal{European Journal of Mechanics - B/Fluids}

\begin{document}

\begin{frontmatter}

\title{Deterministic diffusion models for Lagrangian turbulence: robustness and encoding of extreme events}

\author[label1]{Tianyi Li}
\author[label1,label2]{Flavio Tuteri}
\author[label1]{Michele Buzzicotti}
\author[label1]{Fabio Bonaccorso}
\author[label1]{Luca Biferale}

\affiliation[label1]{organization={Department of Physics and INFN, University of Rome ``Tor Vergata''},
            addressline={Via della Ricerca Scientifica 1}, 
            city={Rome},
            postcode={00133}, 
            country={Italy}}

\affiliation[label2]{organization={Laboratoire de Physique de l’Ecole normale supérieure, ENS, Université PSL, CNRS, Sorbonne Université, Université de Paris},
            addressline={24 Rue Lhomond},
            city={Paris},
            postcode={F-75005},
            country={France}}

% ----- ----- ----- ----- ----- ----- ----- ----- ----- -----  |

\begin{abstract}

Modeling Lagrangian turbulence remains a fundamental challenge due to its multiscale, intermittent, and non-Gaussian nature. Recent advances in data-driven diffusion models have enabled the generation of realistic Lagrangian velocity trajectories that accurately reproduce statistical properties across scales and capture rare extreme events. 
This study investigates three key aspects of diffusion-based modeling for Lagrangian turbulence. First, we assess architectural robustness by comparing a U-Net backbone with a transformer-based alternative, finding strong consistency in generated trajectories, with only minor discrepancies at small scales. Second, leveraging a deterministic variant of diffusion model formulation, namely the deterministic denoising diffusion implicit model (DDIM), we identify structured features in the initial latent noise that align consistently with extreme acceleration events. Third, we explore accelerated generation by reducing the number of diffusion steps, and find that DDIM enables substantial speedups with minimal loss of statistical fidelity. These findings highlight the robustness of diffusion models and their potential for interpretable, scalable modeling of complex turbulent systems.

\end{abstract}

% ----- ----- ----- ----- ----- ----- ----- ----- ----- -----  |


\end{frontmatter}

% ----- ----- ----- ----- ----- ----- ----- ----- ----- -----  |

\section{Introduction}

Understanding the statistical and dynamical properties of Lagrangian turbulence remains a fundamental challenge in fluid dynamics, with implications across atmospheric science, oceanography, and engineering applications~\citep{sawford2001turbulent, yeung2002lagrangian, toschi2009lagrangian}. The Lagrangian viewpoint, which follows individual fluid particles over time, provides key insights into dispersion, intermittency, and extreme event dynamics~\citep{la2001fluid, mordant2001measurement, biferale2004multifractal}. However, despite decades of sustained effort, developing effective models for Lagrangian turbulence remains an open challenge, as turbulence spans a wide range of interacting and non-self-similar time and length scales, from large scales typically dominated by energy injection and characterized by Gaussian statistics, to small scales dominated by dissipation and marked by strong non-Gaussianity and intermittent bursts.

\begin{figure}
    \centering
    \includegraphics[width=\textwidth]{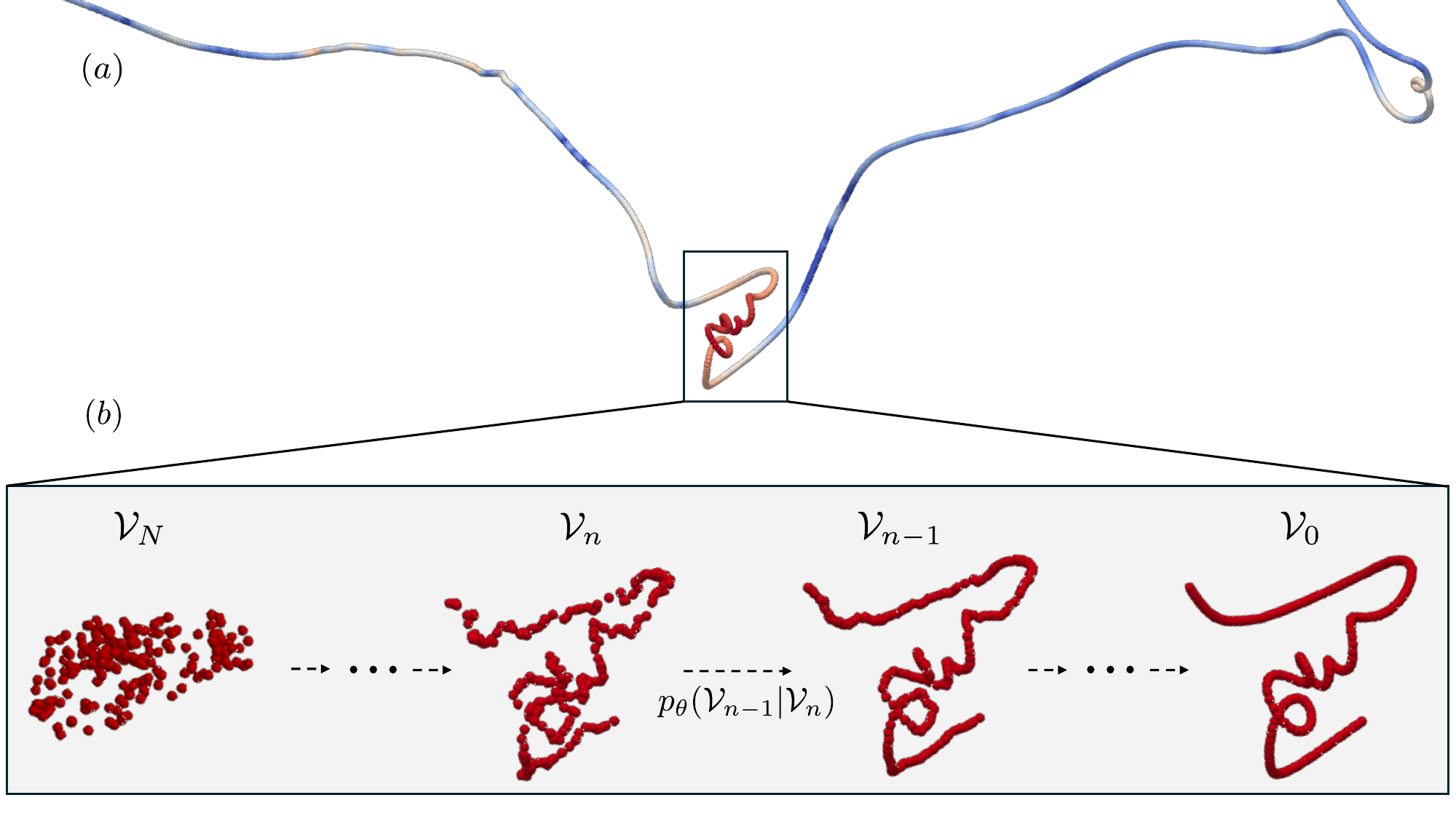}
	\caption{Schematic illustration of the diffusion process. (a) A sample trajectory. (b) From right to left: forward noising process. From left to right: reverse denoising process modeled by a neural network parametrized by $\theta$.}\label{fig:schematic}
\end{figure}

Numerous phenomenological approaches have been proposed, including stochastic models with multiple time scales~\citep{sawford1991reynolds, pope2011simple, viggiano2020modelling}, as well as multifractal and multiplicative cascade-based formulations~\citep{biferale1998mimicking, arneodo1998random, bacry2003log, lubke2023stochastic}. While these models are able to reproduce certain nontrivial features of turbulent statistics, they typically focus on specific regimes and lack the ability to generate synthetic trajectories with accurate multiscale statistics across the full range of turbulent dynamics. In our recent work~\citep{li2024synthetic}, we addressed this limitation through a data-driven approach based on denoising diffusion probabilistic models (DDPMs)~\citep{sohl2015deep, ho2020denoising}. Figure~\ref{fig:schematic}(a) illustrates a typical Lagrangian tracer trajectory generated by a learned denoising diffusion process. Panel (b) of the same figure zooms in on an extreme present in the generated trajectory and illustrates its formation process during denoising diffusion. Trained on high-resolution direct numerical simulation (DNS) data in homogeneous isotropic turbulence, these models can generate Lagrangian velocity trajectories that accurately reproduce high-order statistical properties across a wide range of temporal scales, and provide a practical alternative to data acquisition via DNS or experiments, with substantially reduced computational and experimental overhead. We have demonstrated that this framework can be easily expanded to include tracer, light, and heavy inertial particles while maintaining strong agreement with reference statistics~\citep{li2024generative}. More recently, we have also shown how to condition the generation to solve the reconstruction problem~\citep{buzzicotti2023data} when only gappy Lagrangian data is available~\citep{li2024stochastic}.

Despite these advances, several important questions remain open. First, the extent to which diffusion model performance depends on neural network architecture has not been systematically evaluated. This question is particularly important in physical settings, where architectural robustness provides insight into whether the learned generative process reflects genuine physical dynamics or is overly sensitive to implementation details. Most existing diffusion models employ a convolutional U-Net backbone~\citep{ronneberger2015u}, which has been the standard architecture in image synthesis since the seminal work of~\citet{ho2020denoising}, and remains the dominant choice in subsequent developments~\citep{nichol2021improved, dhariwal2021diffusion}. Our previous studies on synthetic Lagrangian turbulence also adopted a U-Net architecture~\citep{li2024synthetic, li2024generative, martin2025generation}. More recently, Diffusion Transformers (DiTs)~\citep{peebles2023scalable}, built on the best practices of Vision Transformers (ViTs)~\citep{dosovitskiy2020image}, have demonstrated that the U-Net backbone can be effectively replaced by a transformer in image generation tasks. Transformers offer practical advantages over U-Nets, including greater scalability and more systematic control over model capacity. These properties make transformers a promising alternative for future applications involving larger-scale and higher-Reynolds-number Lagrangian turbulence.

Second, while our previous work has shown that diffusion models can reproduce and generalize rare and intermittent events with high statistical fidelity in both the Eulerian~\citep{li2023multi} and Lagrangian~\citep{li2024synthetic} frames, the mechanism by which such extreme fluctuations arise during generation remains unclear. We now turn to a more focused question: can we empirically understand how such events are constructed within the diffusion framework? In DDPM, generation proceeds through a sequence of stochastic transitions, with new noise injected at each step. As a result, the output reflects the cumulative influence of both the initial latent and the per-step noise, making it challenging to attribute specific features, such as extreme events, to individual sources. In contrast, the Denoising Diffusion Implicit Model (DDIM)~\citep{song2020denoising} defines a deterministic variant of DDPM, where the output trajectory is fully determined by the initial input noise. This makes it possible to explore whether there exists a systematic connection between extreme events and structured fluctuations in the latent input. Such analysis requires first verifying that DDIM retains statistical fidelity comparable to DDPM.

Third, the standard DDPM framework requires hundreds to thousands of iterative denoising steps to generate each trajectory, which can limit its practical applicability in large-scale or real-time scenarios. Recent work in image generation~\citep{song2020denoising, nichol2021improved} has shown that the number of sampling steps can be substantially reduced at inference time for both DDPM and DDIM, enabling significant acceleration without retraining. Whether such step-reduction strategies can be effectively applied in the context of Lagrangian turbulence, without compromising the fidelity of multiscale statistics, remains an open and practically important question.

The rest of this paper is organized as follows. Section~\ref{sec:methodology} discusses the dataset, a unified generative framework encompassing DDPM and DDIM, the accelerated generation strategy, the network architecture, and the training details. Section~\ref{sec:results} presents our main findings on model robustness across architectures, the latent signatures of extreme events under DDIM, and the performance of step-reduced generation, with both DDPM and DDIM sampling schemes used where applicable. Section~\ref{sec:conclusions} summarizes our findings and outlines directions for future research.

% ----- ----- ----- ----- ----- ----- ----- ----- ----- -----  |

\section{Methodology}\label{sec:methodology}

\subsection{Lagrangian Turbulence Dataset}

In this study, we use the same dataset of Lagrangian tracer trajectories as in our previous work~\citep{li2024synthetic}. The trajectories are obtained by tracking passive point-like particles in a direct numerical simulation (DNS) of three-dimensional incompressible turbulence, conducted in a cubic periodic domain with a grid resolution of $1024^3$. The Eulerian velocity field is computed by solving the Navier–Stokes equations using a fully dealiased pseudo-spectral method with large-scale isotropic forcing, reaching a statistically stationary state with a Taylor-scale Reynolds number of $R_\lambda \approx 310$. Details of the simulation setup, along with key Eulerian and Lagrangian statistics, can be found in~\citep{biferale2023turb, calascibetta2023optimal}.

Once statistical stationarity is achieved, $N_p = 327{,}680$ passive tracers are randomly seeded in the domain and advected according to $\bm{V}(t) = \dot{\bm{X}}(t) = \bm{u}(\bm{X}(t), t)$, where $\bm{X}(t)$ and $\bm{V}(t)$ denote the particle position and velocity at time $t$, respectively, and $\bm{u}$ is the Eulerian velocity field. The particle motion is integrated numerically using sixth-order B-spline interpolation for velocity evaluation and a second-order Adams–Bashforth method for time integration. Velocity data are recorded at regular intervals $\Delta t \simeq 0.1\tau_\eta$, where $\tau_\eta$ is the Kolmogorov time scale, over a total duration of $T \simeq 1.3\tau_L \simeq 200\tau_\eta$, with $\tau_L$ the large-eddy turnover time. Each trajectory is thus discretized into $K = 2000$ time steps, and represented as
\begin{equation}\label{eq:traj}
    \CV = \{ V_x(t_k), V_y(t_k), V_z(t_k) \mid t_k \in [0, T];\, k = 1,\dots,K \},
\end{equation}
where $V_i(t_k)$ is the $i$-th component of the particle velocity at time $t_k$.

% ----- ----- ----- ----- ----- ----- ----- ----- ----- -----  |

\subsection{A Broad Class of Generative Processes: From DDPM to DDIM}\label{sec:ddpm_to_ddim}

Our objective is to model the data distribution $q(\CV)$ of the ground-truth trajectories defined in Eq.~\eqref{eq:traj}, by constructing a forward noising process and learning the corresponding reverse denoising process via a neural network. The forward process progressively perturbs a clean trajectory $\CV \sim q(\CV)$, drawn from the training data, over $N$ steps by adding Gaussian noise at each step. We denote the initial trajectory as $\CV_0 \coloneqq \CV$, and let $\CV_{1:N} \coloneqq \{\CV_1, \CV_2, \dots, \CV_N\}$ denotes the full sequence of noisy states.

\begin{figure}
    \centering
    \includegraphics[width=0.8\textwidth]{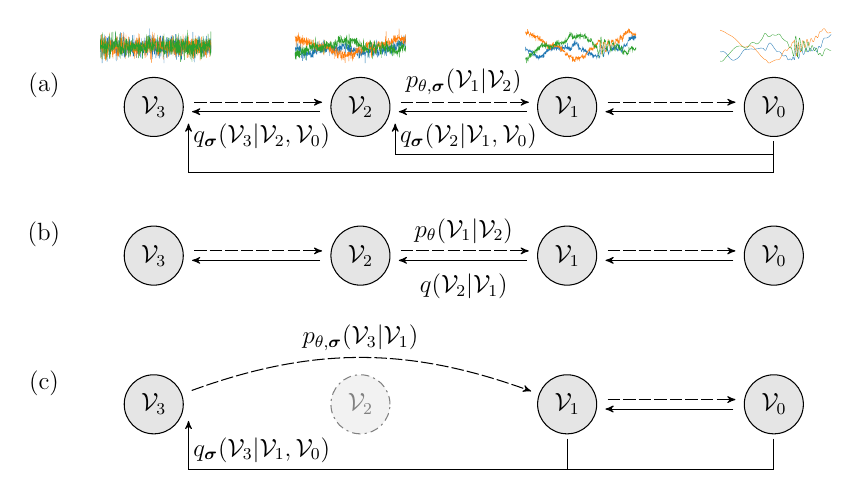}
	\caption{Graphical illustrations of the diffusion frameworks with a small number of steps ($N = 3$) shown for ease of illustration. Solid arrows represent the forward process, while dashed arrows indicate the reverse process modeled by a neural network $p_{\theta,\bm{\sigma}}(\CV_{n-1}|\CV_n)$. (a) General diffusion with a non-Markovian forward process $q_{\bm{\sigma}}(\CV_n|\CV_{n-1}, \CV_0)$, where each step depends on both $\CV_{n-1}$ and $\CV_0$, while preserving the marginal distribution $q(\CV_n|\CV_0)$. (b) DDPM: a Markovian forward process $q(\CV_n|\CV_{n-1})$ progressively adds Gaussian noise to the clean trajectory $\CV_0$. The reverse process denoises step by step from $\CV_N$ back to $\CV_0$. (c) Accelerated generation using a subset of $M = 2$ steps, with index set $\mathcal{S} = \{1, 3\}$ indicating the retained steps.}\label{fig:ddpm_ddim_subset}
\end{figure}

We are particularly interested in a class of forward processes that share the same Gaussian marginal distribution at each step $n$. These marginals are fully determined by a predefined noise schedule $\bm{\bar{\alpha}} = \{\bar{\alpha}_n\}_{n=1}^N$, and take the form:
\begin{equation}
\label{eq:marginal}
q_{\bm{\bar{\alpha}}}(\mathcal{V}_n|\mathcal{V}_0) = \mathcal{N}(\sqrt{\bar{\alpha}_n} \mathcal{V}_0, (1 - \bar{\alpha}_n)\bm{I}) \,.
\end{equation}
We omit the subscript $\bm{\bar{\alpha}}$ in what follows for clarity, as the schedule is fixed throughout. The schedule is typically chosen such that $\bar{\alpha}_1 \approx 1$ and $\bar{\alpha}_N = 0$, inducing a near-continuous transformation from the data distribution $q(\CV_0)$ to a standard Gaussian distribution, $q(\CV_N) = \mathcal{N}(\mathbf{0}, \bm{I})$. The corresponding family of forward processes, indexed by parameters $\bm{\sigma} = \{\sigma_n\}_{n=1}^N$, is defined by:
\begin{equation}
\label{eq:fchain}
q_{\bm{\sigma}}(\CV_{1:N}|\CV_0) \coloneqq \prod_{n=1}^N q_{\bm{\sigma}}(\CV_n|\CV_{n-1}, \CV_0) \,,
\end{equation}
where each transition step $q_{\bm{\sigma}}(\CV_n|\CV_{n-1}, \CV_0)$ depends on both the previous state $\CV_{n-1}$ and the original trajectory $\CV_0$, as illustrated in Fig.~\ref{fig:ddpm_ddim_subset}(a).

We now define the form of each transition distribution $q_{\bm{\sigma}}(\CV_n|\CV_{n-1}, \CV_0)$. Each transition distribution is assumed to be Gaussian, such that the product in Eq.~\eqref{eq:fchain} yields Gaussian marginals, as requested by Eq.~\eqref{eq:marginal}. Using Bayes' theorem, the reverse transition is given by:
\begin{equation}
\label{eq:bayes}
q_{\bm{\sigma}}(\CV_{n-1}|\CV_n, \CV_0)
= \frac{q_{\bm{\sigma}}(\CV_n|\CV_{n-1}, \CV_0) \cdot q(\CV_{n-1}|\CV_0)}{q(\CV_n|\CV_0)} \,.
\end{equation}
Since all terms on the right-hand side are Gaussian, the reverse transition is also Gaussian. We therefore consider the
class of models parametrized as:
\begin{equation}
\label{eq:general_back}
q_{\bm{\sigma}}(\CV_{n-1}|\CV_n, \CV_0) = \mathcal{N}(\omega_n \CV_n + \rho_n \CV_0, \sigma_n^2 \bm{I}) \,,
\end{equation}
where the indexing parameter $\sigma_n$ determines the variance of the Gaussian reverse transition distribution. Its mean is a linear combination of $\CV_n$ and $\CV_0$. The coefficients $\omega_n$ and $\rho_n$ are derived by combining Eq.~\eqref{eq:general_back} and  Eq.~\eqref{eq:marginal}, see~\ref{app:app1}, and result in,
\begin{equation}
\label{eq:sol}
\omega_n = \sqrt{\frac{1 - \bar{\alpha}_{n-1} - \sigma_n^2}{1 - \bar{\alpha}_n}}, \quad
\rho_n = \sqrt{\bar{\alpha}_{n-1}} - \sqrt{\bar{\alpha}_n} \, \omega_n \,.
\end{equation}
The corresponding forward transition distribution can be explicitly written as,
\begin{align}
\label{eq:general_forw}
& q_{\bm{\sigma}}(\CV_n|\CV_{n-1}, \CV_0) 
= \nonumber \\
& \mathcal{N} \left(
\frac{1}{1 - \bar{\alpha}_{n-1}} \left( \sqrt{\bar{\alpha}_n} \sigma_n^2 \CV_0 
+ \omega_n (1 - \bar{\alpha}_n)(\CV_{n-1} - \rho_n \CV_0) \right), 
\frac{1 - \bar{\alpha}_n}{1 - \bar{\alpha}_{n-1}} \sigma_n^2 \bm{I}
\right).
\end{align}

The goal of diffusion models is to approximate the generalized reverse distribution, defined in Eq.~\eqref{eq:general_back} with Eq.~\eqref{eq:sol}, without knowing $\CV_0$, but using only $\CV_n$.
That is, each generalized backward step is parameterized by a neural network with trainable parameters $\theta$, such that $p_{\theta,\bm{\sigma}}(\CV_{n-1}|\CV_n) \approx q_{\bm{\sigma}}(\CV_{n-1}|\CV_n, \CV_0)$. Once trained, as will be discussed below, the generative model starts at step $N$ from Gaussian noise, $\CV_N \sim q(\CV_N)=\mathcal{N}(\bm{0}, \bm{I})$, and iteratively produces $\CV_{n-1}$ from $\CV_n$ to $\CV_0$. The full generalized generative process is defined as
\begin{equation}\label{eq:gen_process}
    p_{\theta,\bm{\sigma}}(\CV_{0:N}) = q(\CV_N) \prod_{n=1}^N p_{\theta,\bm{\sigma}}(\CV_{n-1}|\CV_n).
\end{equation}
To accomplish this goal, the neural network needs to learn how to estimate $\CV_0$ from the knowledge of its noisy representation, $\CV_n$. 
From Eq.~\eqref{eq:marginal} we known that each noisy sample $\CV_n$ is related to $\CV_0$ by the following simple relation, also known as reparameterization trick:
\begin{equation}
\label{eq:forward_reparam}
\CV_n = \sqrt{\bar{\alpha}_n} \CV_0 + \sqrt{1 - \bar{\alpha}_n} \, \bm{\epsilon}, \quad \bm{\epsilon} \sim \mathcal{N}(\bm{0}, \bm{I}).
\end{equation}
It follows that if the neural network is able to extract the noise term in $\CV_n$, namely $\bm{\epsilon}_\theta(\CV_n,n)\approx \bm{\epsilon}$, it can get an approximation of $\CV_0$ by inverting Eq.~\eqref{eq:forward_reparam} as follows,
\begin{equation}
\widehat{\CV}_{0,\theta} := \frac{1}{\sqrt{\bar{\alpha}_n}} \left( \CV_n - \sqrt{1 - \bar{\alpha}_n} \, \bm{\epsilon}_\theta(\CV_n, n) \right).
\end{equation}
In this way, the posterior of the forward process can be modeled as
\begin{equation}
\label{eq:reverse_theta}
p_{\theta,\bm{\sigma}}(\CV_{n-1}|\CV_n) \coloneqq
\mathcal{N}\left(
\omega_n \CV_n + \rho_n \widehat{\CV}_{0,\theta},
\sigma_n^2 \bm{I}
\right) \approx \mathcal{N}\left(
\omega_n \CV_n + \rho_n \CV_0,
\sigma_n^2 \bm{I}
\right),
\end{equation}
where $\omega_n$ and $\rho_n$ are always the same as in Eq.~\eqref{eq:sol}.
The neural network is trained to minimize the negative log-likelihood:
\begin{equation}\label{equ:nll}
    \mathbb{E}_{q(\CV_0)}[-\log(p_{\theta,\bm{\sigma}}(\CV_0))],
\end{equation}
which is estimated through a tractable upper bound. This leads to a simplified mean squared error loss that is independent of the variance parameters, $\bm{\sigma}$ ~\citep{song2020denoising},

\begin{equation}\label{equ:L_simple}
    L_\mathrm{simple} = \mathbb{E}_{n,\,q(\CV_0),\,\bm{\epsilon}} \left[ \left\| \bm{\epsilon} - \bm{\epsilon}_\theta\left(\CV_n(\CV_0, \bm{\epsilon}), n \right) \right\|^2 \right],
\end{equation}
where $\CV_n(\CV_0, \bm{\epsilon})$ is generated from the clean sample $\CV_0$ and Gaussian noise $\bm{\epsilon}$ via the forward reparameterization in Eq.~\eqref{eq:forward_reparam}. Further discussion about the training procedure can be found in~\citep{ho2020denoising,li2024synthetic}.

DDPM and DDIM arise as special cases within this generalized process family. To get the DDPM we need to set the parameters $\sigma_n$ such that to have a Markovian forward process~\citep{ho2020denoising}. It follows,
\begin{equation}
    \sigma_n^2 = \frac{1 - \bar{\alpha}_{n-1}}{1 - \bar{\alpha}_n} \left(1 - \frac{\bar{\alpha}_n}{\bar{\alpha}_{n-1}} \right), \quad \text{with } \bar{\alpha}_0 \coloneqq 1.
\end{equation}
DDIM is another special case that arises in the zero-variance limit $\sigma_n \to 0$ for all $n$, resulting in a backward procedure that maps the initial Gaussian noise $\CV_N$ to a synthetic trajectory $\CV_0$ through a sequence of deterministic transformations. Thus in DDIM, the joint distribution in Eq.~\eqref{eq:gen_process} is no longer a valid density, and the model becomes implicitly probabilistic~\citep{song2020denoising}.

Since training is independent of the choice of $\bm{\sigma}$, the same neural network trained to predict $\bm{\epsilon}_\theta(\CV_n,n)$ can be used to model any of the generalized backward processes. This reuse also applies when generation is performed on a reduced subset of diffusion steps, as discussed in the next section.

% ----- ----- ----- ----- ----- ----- ----- ----- ----- -----  |

\subsection{Accelerated Generation via Subset Diffusion Steps}\label{sec:step_reduction}

The generative process, in both DDPM and DDIM formalisms, consists of $N$ iterative steps, sequentially sampling each intermediate state from $\CV_N$ down to $\CV_0$ by evaluating the neural network at each step. As the computational cost scales linearly with $N$, this motivates reducing the number of steps used during sampling to accelerate generation.

To this end, we define a reduced generative process that retains the exact formulation introduced in Section~\ref{sec:ddpm_to_ddim}, but operates over a selected subset of diffusion steps from the original process. Specifically, the new process consists of $M < N$ steps, with a noise schedule $\{\bar{\alpha}_{s_i}\}_{i=1}^M$ extracted from the original schedule $\{\bar{\alpha}_n\}_{n=1}^N$. The index set $\mathcal{S} = \{s_1, \dots, s_M\} \subseteq \{1, \dots, N\}$ specifies an increasing sequence of selected diffusion steps (see Fig.~\ref{fig:ddpm_ddim_subset}(c) for a schematic example).

When $M$ is much smaller than $N$, this reduction significantly improves sampling efficiency by reducing the number of network evaluations. Importantly, \citet{song2020denoising} justified that the noise prediction network $\bm{\epsilon}_\theta$, trained on the full diffusion process under the DDPM objective Eq.~\eqref{equ:L_simple}, remains optimal for the reduced process. This enables flexible trade-offs between generation speed and fidelity by sampling with different numbers of steps using a single pretrained model.

\subsection{Network Architectures and Training Setup}

We compare two representative architectures for the noise prediction network $\bm{\epsilon}_\theta(\CV_n, n)$: a convolutional U-Net and a transformer-based architecture. The U-Net architecture is exactly the same as in our previous work~\citep{li2024synthetic}, to which we refer the reader for full details (see Figure 3(a) and the Methods section therein).

\begin{figure}
    \begin{tikzpicture}
        \node [minimum width=0.45\textwidth, align=center] (archfig) {
            \includegraphics[width=0.45\textwidth]{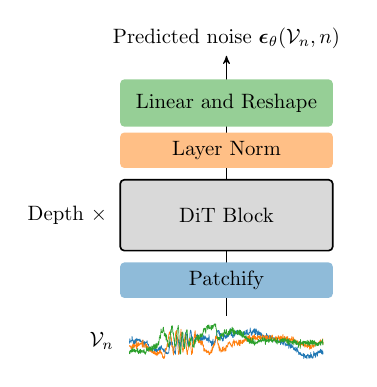}
        };
        \node [minimum width=0.45\textwidth, align=center, right=0 of archfig] (hparam) {
            \small
            Hyperparameters \\[0.5em]
            \small
            \begin{tabular}{|l|c|}
                \hline
                Diffusion steps & 800 \\ \hline
                Noise Schedule & tan6-1 \\ \hline
                Model size & 61M \\ \hline
                Depth & 12 \\ \hline
                Patch size & 8 \\ \hline
                Hidden size & 528 \\ \hline
                Num heads &  6 \\ \hline
                Batch size & 256 \\ \hline
                Learning Rate & 1e-4 \\ \hline
            \end{tabular}
        };
    \end{tikzpicture}
    \caption{Overview of the DiT-based architecture (\textit{left}) and associated architectural and training hyperparameters (\textit{right}). ``Depth'' indicates the number of transformer blocks, and ``Num heads'' the number of attention heads per block. See main text for definitions of all other parameters.}\label{fig:dit_overview}
\end{figure}

For the transformer-based architecture, we adopt the best-performing DiT configuration as presented in~\citet{peebles2023scalable}. A schematic overview is shown in Fig.~\ref{fig:dit_overview} (left). We introduce only two minimal modifications. First, the patchify layer is adapted to process noised trajectories of shape $(K, 3)$ by dividing them along the temporal axis into $K/p$ non-overlapping patches (where $p$ is the patch size), each of which is linearly embedded into a token with dimension equal to the hidden size. Second, the network is configured to predict only the denoised noise $\bm{\epsilon}_\theta(\CV_n, n)$, without producing a covariance output. All other architectural components remain unchanged. The core of the model consists of multiple transformer blocks (DiT blocks) using the AdaLN-Zero variant to incorporate diffusion step conditioning~\citep{peebles2023scalable}.

We train both architectures under the same conditions for direct comparability. Specifically, we use 800 diffusion steps with a tan6-1 noise schedule~\citep{li2024synthetic}, a batch size of 256, and a fixed learning rate of $10^{-4}$ with the AdamW optimizer~\citep{loshchilov2017decoupled}. Both models are trained for $4 \times 10^5$ iterations, and an exponential moving average (EMA) with a decay rate of 0.999 is maintained during training and used at inference time. The transformer-based model (DiT) matches the U-Net in parameter count, with no attempt to optimize the architecture. Its architectural and training hyperparameters are summarized in Fig.~\ref{fig:dit_overview} (right). Training is performed on 4 NVIDIA A100 GPUs and takes approximately 38 hours for each model.

\section{Results and Discussion}\label{sec:results}

\subsection{Architectural Robustness of Diffusion Models}

To evaluate the architectural robustness of diffusion models, we consider three representative configurations: U-Net with DDPM (UN-P), U-Net with DDIM (UN-I), and Transformer with DDIM (TF-I). While our primary focus is on architectural effects, we also vary the diffusion scheme---from the stochastic DDPM to the deterministic DDIM---as an additional probe of robustness in reproducing the multiscale statistics of Lagrangian turbulence.

We focus on three statistical measures that capture the multiscale behavior of Lagrangian turbulence. The first is the $p$-th order Lagrangian structure function,
\begin{equation}
    S_\tau^{(p)} = \langle [V_i(t+\tau) - V_i(t)]^p \rangle,
\end{equation}
where $\tau$ denotes the temporal separation scale of interest. The angle brackets indicate averaging over time and across particle trajectories. Here, $i = x, y, z$ denotes the velocity components, and we omit this index in $S^{(p)}_\tau$ under the assumption of isotropy. The second quantity is the generalized flatness,
\begin{equation}
    F_\tau^{(p)} = \frac{S_\tau^{(p)}}{[S_\tau^{(2)}]^{p/2}},
\end{equation}
which characterizes scale-dependent intermittency. For Gaussian-distributed velocity increments, $F^{(4)}_\tau = 3$, while larger values reflect increasingly heavy-tailed, intermittent statistics. Finally, we consider the local scaling exponent from extended self-similarity (ESS)~\citep{benzi1993extended, arneodo2008universal},
\begin{equation}
    \zeta(p,\tau) = \frac{d \log S_\tau^{(p)}}{d \log S_\tau^{(2)}},
\end{equation}
which serves as a stringent and quantitative multiscale benchmark. Unlike the structure function or flatness, which vary significantly across scales, $\zeta(p,\tau)$ remains an $\mathcal{O}(1)$ quantity  across multiple decades of time lags, enabling high-precision assessment of multiscale statistical behavior.

\begin{figure}
  \includegraphics[width=\textwidth]{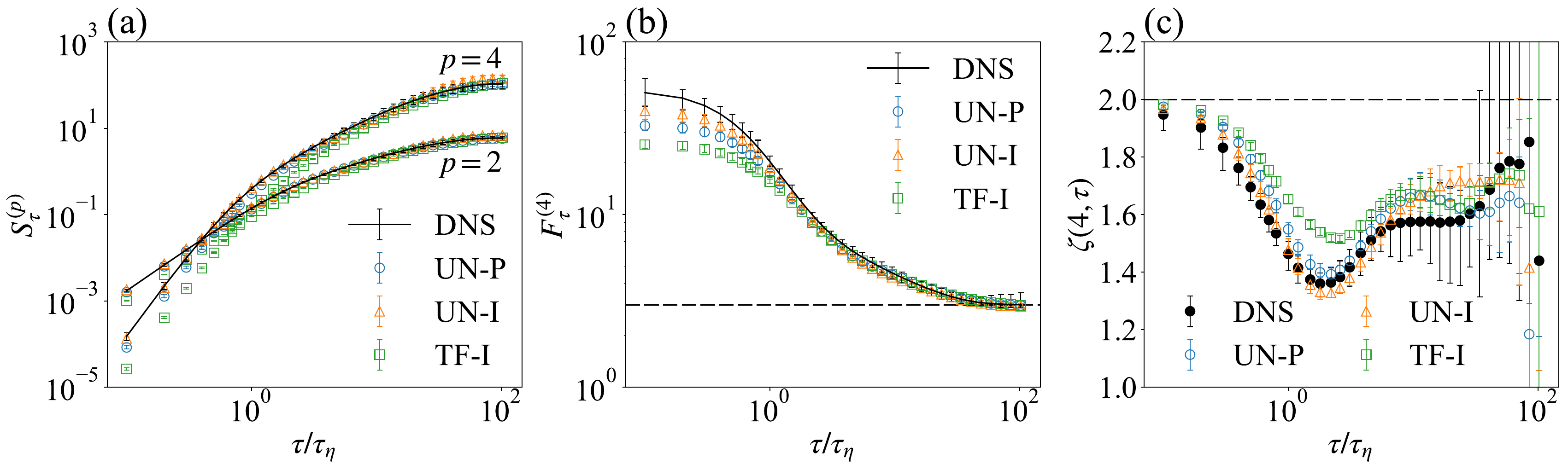}
    \caption{Comparison of Lagrangian statistics generated by different model architectures and diffusion schemes. Results are shown for three configurations: U-Net with DDPM (UN-P), U-Net with DDIM (UN-I), and Transformer with DDIM (TF-I). The black solid line corresponds to the DNS reference. (a) Log-log plots of Lagrangian structure functions $S^{(p)}_\tau$ for $p=2,4$; (b) Fourth-order generalized flatness $F^{(4)}_\tau$. The horizontal dashed line at $F^{(4)}_\tau=3$ corresponds to Gaussian velocity increments. (c) Fourth-order logarithmic local slope $\zeta(4,\tau)$. The horizontal dashed line indicates the non-intermittent dimensional scaling $\zeta(4) = 2$, i.e., $S^{(4)}_\tau \propto [S^{(2)}_\tau]^2$. Mean and error bars are computed across 30 batches derived from $N_p$ trajectories, with 10 batches per velocity component; error bars indicate the full min–-max range across batches.}\label{fig:model_comparison}
\end{figure}

Fig.~\ref{fig:model_comparison} summarizes the multiscale statistical performance of the three model configurations. Across all three diagnostics---structure functions, flatness, and local slopes---both UN-P and UN-I show excellent agreement with the DNS reference over the entire range of time lags. The transformer model with DDPM (TF-P, not shown) also performs well at intermediate and large scales, but tends to underestimate intermittency at small scales, with lower $F^{(4)}_\tau$ values and a shallower dip in $\zeta(4,\tau)$ when $\tau/\tau_\eta \lesssim 2$. This underestimation becomes more pronounced in the local slope $\zeta(4,\tau)$ when switching to deterministic sampling in TF-I, which exhibits further degradation at small scales, while still maintaining reasonable accuracy at larger scales.

\begin{figure}
    \includegraphics[width=\textwidth]{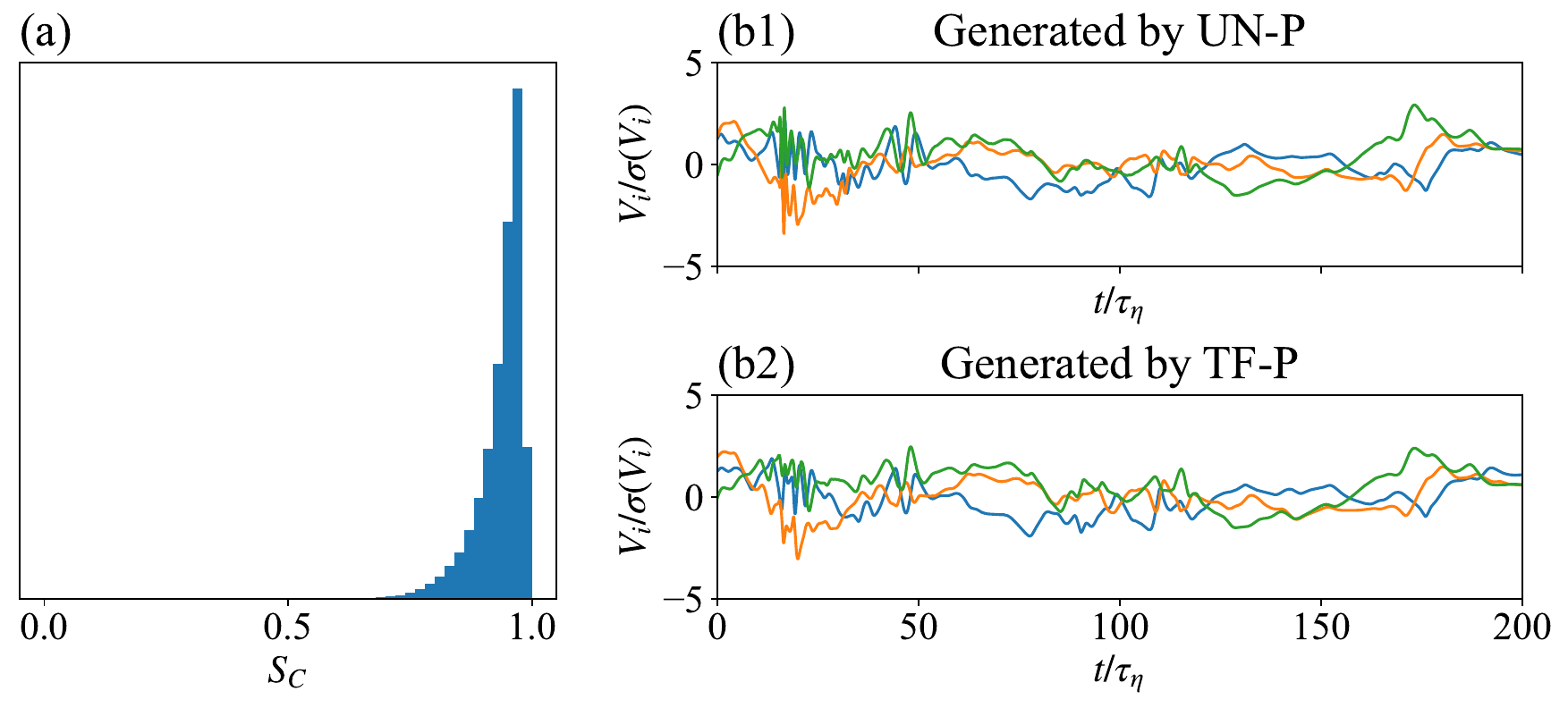}
    \caption{Comparison of generations from UN-P and TF-P using identical random sequences in the backward diffusion process. (a) Distribution of cosine similarity between outputs of the two models, showing a sharp peak near 1.0, indicating strong agreement across architectures. (b) A representative trajectory pair, showing strong overall similarity, with slightly reduced small-scale fluctuations in TF-P around $t/\tau_\eta \approx 20$. Different colors correspond to different velocity components.}\label{fig:trajectory_correlation}
\end{figure}

Despite the small-scale differences observed in statistical diagnostics, we further assess whether the two architectures produce consistent trajectory-level behavior. To this end, we generate trajectories from UN-P and TF-P using \textit{identical random sequences} (i.e., the full sequence of sampling noise) throughout the reverse diffusion process in Eq.~\eqref{eq:gen_process}, thereby eliminating stochastic variability.

To quantify the alignment between the two models’ outputs, we compute the cosine similarity between corresponding pairs of velocity trajectories generated with the same randomness:
\begin{equation}
    S_C=\frac{\int V_i^{\mathrm{(UN\text{-}P)}}V_i^{\mathrm{(TF\text{-}P)}}\,dt}{(\int[V_i^{\mathrm{(UN\text{-}P)}}]^2\,dt)^{1/2}(\int[V_i^{\mathrm{(TF\text{-}P)}}]^2\,dt)^{1/2}},
\end{equation}
where $V_i^{\mathrm{(UN\text{-}P)}}$ and $V_i^{\mathrm{(TF\text{-}P)}}$ denote the $i$-th velocity components of a pair of trajectories generated by the two models. Summation over $i$ is implied, and the integral is taken over the full temporal extent of each trajectory, from $0$ to $T$.

The distribution of cosine similarity values computed over $N_p$ trajectory pairs is shown in Fig.~\ref{fig:trajectory_correlation}(a). The strong peak near 1.0 demonstrates that the two architectures produce highly consistent outputs under identical sampling conditions.

A representative trajectory pair is shown in Fig.~\ref{fig:trajectory_correlation}(b), further illustrating this agreement: both trajectories exhibit nearly identical large- and intermediate-scale structures, with TF-P showing slightly reduced small-scale fluctuations around $t/\tau_\eta \approx 20$, consistent with the underestimation of small-scale intermittency observed in the statistical diagnostics. This behavior underscores the ability of diffusion models to encode a shared representation of the underlying physical process, despite architectural differences. This may reflect the benefit of the diffusion framework’s inductive bias, as also suggested in recent theoretical work~\citep{kadkhodaie2023generalization}.

Together, these results indicate that the diffusion model framework promotes robustness across architectures at most physical scales, while small-scale accuracy may depend more sensitively on the choice of network structure. We emphasize that no architectural tuning was performed for the transformer model, suggesting that further optimization could significantly enhance its small-scale performance.

\subsection{Latent Noise Signatures of Extreme Events under DDIM}

Extreme events---such as sharp bursts of acceleration---are rare but physically significant features of Lagrangian turbulence. These events often reside in the far tails of the acceleration distribution, reaching several tens of standard deviations. Having assessed the robustness of diffusion models across both architectures and sampling schemes, we now leverage the deterministic nature of DDIM in the UN-I model to investigate whether rare, high-acceleration events in generated trajectories can be systematically traced back to structured patterns in the initial latent noise.

\begin{figure}
    \centering
    \includegraphics[width=\textwidth]{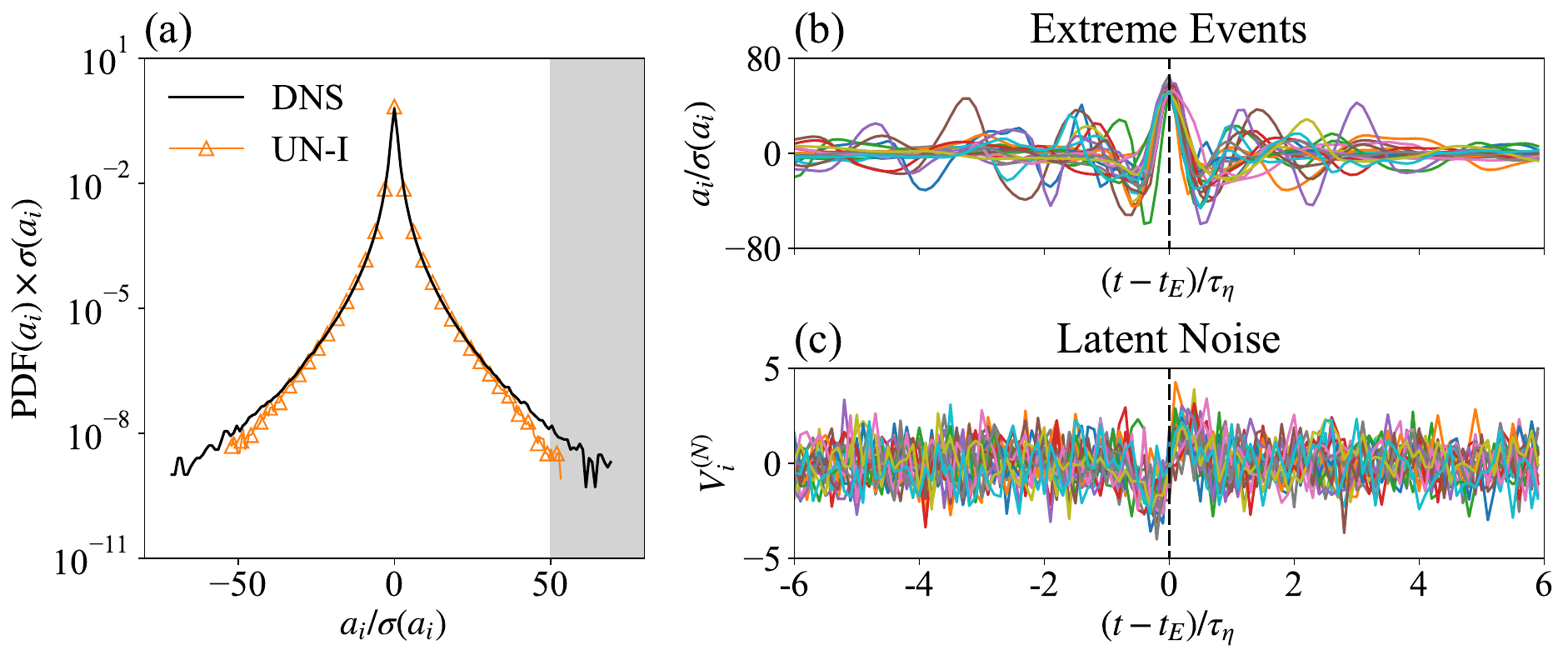}
    \caption{Analysis of extreme acceleration events and their latent noise signatures under DDIM sampling with the U-Net backbone (UN-I). (a) Standardized PDFs of acceleration $a_i$, aggregated over all velocity components, for DNS reference data and synthetic trajectories generated by DDIM. Acceleration values are normalized by the standard deviation $\sigma(a_i)$ of the DNS data. The gray shaded region highlights extreme events with $a_i / \sigma(a_i) \geq 50$, corresponding to large positive acceleration excursions. (b) Acceleration profiles aligned at the time $t_E$ of maximum positive excursion in each selected trajectory from the gray region in (a). For each trajectory, only the component with the largest $a_i$ is retained and centered such that $t - t_E = 0$. (c) Corresponding latent noise inputs $V_i^{(N)}$, sampled from the initial Gaussian distribution and used by DDIM to generate the trajectories in (b). The same component and alignment convention are applied.}\label{fig:extreme_events}
\end{figure}

Fig.~\ref{fig:extreme_events}(a) compares the probability density function (PDF) of acceleration components $a_i = dV_i/dt$ between DNS data and synthetic trajectories generated by UN-I. The two distributions closely match, including the far tails where extreme events occur. To examine whether such events are linked to patterns in the DDIM input noise, we focus on large positive acceleration excursions, selecting samples with $a_i / \sigma(a_i) \geq 50$, as indicated by the shaded region in Fig.~\ref{fig:extreme_events}(a). For each selected trajectory, we identify the acceleration component and the time $t_E$ at which the maximum of $a_i$ occurs. We then shift this peak to $t - t_E = 0$ and retain only the corresponding component. The aligned acceleration profiles for the selected component are shown in Fig.~\ref{fig:extreme_events}(b). Panel (c) displays the corresponding initial latent noise vectors, aligned using the same procedure and component as in panel (b).

The profiles in Fig.~\ref{fig:extreme_events}(c) reveal a consistent localized increase in the input noise near the origin, mirroring the alignment of acceleration spikes in panel (b). This visual correspondence indicates that extreme acceleration events tend to be associated with structured fluctuations in the latent input, which is sampled from a standard Gaussian distribution.

This empirical correspondence---emerging despite the high dimensionality and randomness of the latent space---suggests that rare physical phenomena may leave discernible signatures in the generative input. Such findings could inform future efforts toward controlled trajectory generation, targeted sampling of extreme events, or deeper interpretability of learned representations in physics-based generative models.

\subsection{Consistency and Sensitivity in Step-Reduced Sampling}

To assess the effect of step reduction on sampling efficiency and statistical accuracy, we apply the subset-based reverse diffusion schedules described in Section~\ref{sec:step_reduction} to both DDPM and DDIM. This strategy, introduced in prior work on image generation~\citep{song2020denoising, nichol2021improved}, enables significantly faster sampling with limited quality loss. We now examine whether such acceleration remains effective in the context of Lagrangian turbulence generation, where preserving physical realism across scales is critical. Specifically, let $\mathcal{S} = \{s_1, \dots, s_M\} \subseteq \{1, \dots, N\}$ denote a monotonic subset of $M$ reverse diffusion steps selected from the full set of $N = 800$ steps. Following the DDIM paper~\citep{song2020denoising}, we adopt a uniform stride schedule defined by
\begin{equation}
    s_i = 1 + \frac{N}{M}(i - 1),
\end{equation}
where $M$ is chosen such that $N/M$ is an integer. This schedule is applied identically to both DDPM and DDIM. We also tested the alternative diffusion step selection proposed in~\citep{nichol2021improved}, which samples $M$ evenly spaced real-valued steps between $1$ and $N$ (inclusive) and rounds them to integers. In our setting, this produced slightly worse results for DDPM and noticeably degraded the performance of DDIM.

\begin{figure}
    \includegraphics[width=\textwidth]{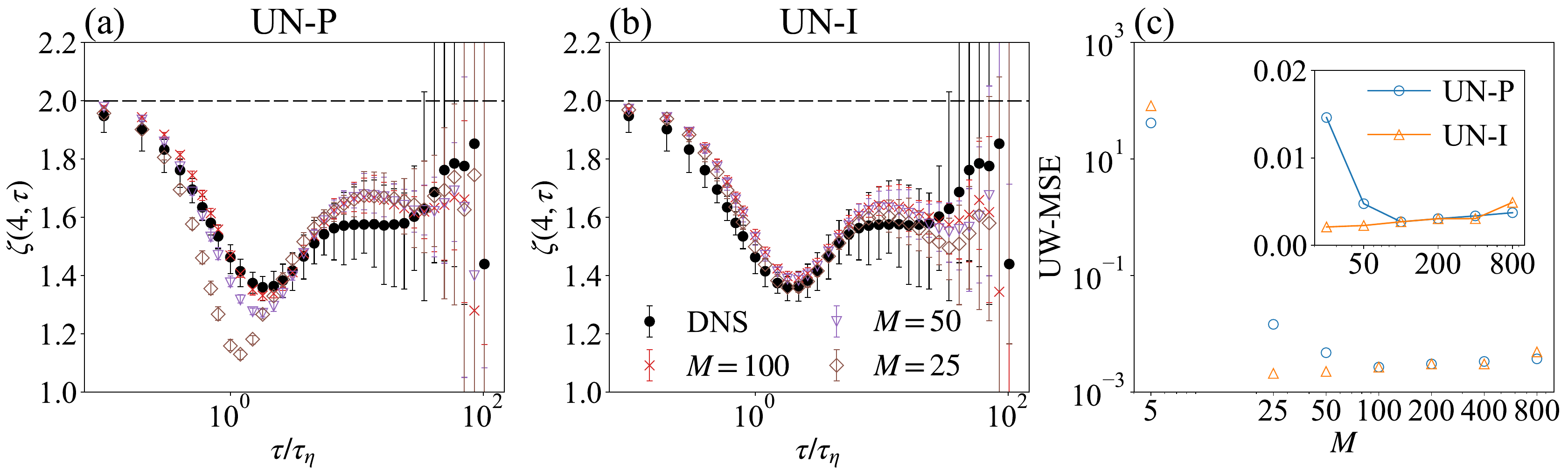}
    \caption{Multiscale statistical behavior under reduced-step sampling for (a) DDPM and (b) DDIM, both using the U-Net backbone. Each panel shows the fourth-order local slope $\zeta(4,\tau)$ for different numbers of reverse diffusion steps $M$ selected from a total of $N = 800$ steps. The horizontal dashed line marks the non-intermittent dimensional scaling, $S^{(4)}_\tau \propto [S^{(2)}_\tau]^2$. Panel (c) reports the uncertainty-weighted MSE (UW-MSE) between generated and DNS-based $\zeta(4,\tau)$ as a function of $M$. Mean and error bars in (a) and (b) are computed from 30 batches (10 per velocity component) over $N_p$ total trajectories; error bars indicate the full min--max range. Legend shared between (a) and (b).}\label{fig:step_reduction}
\end{figure}

Figs.~\ref{fig:step_reduction}(a) and (b) show the fourth-order local slope $\zeta(4,\tau)$ computed from synthetic trajectories generated by UN-P (DDPM) and UN-I (DDIM), respectively, using step counts $M = 100$, 50, and 25. At $M = 100$, both models show close agreement with the DNS reference across scales. As $M$ decreases, DDPM begins to exhibit noticeable degradation, particularly at small scales, while DDIM remains consistently accurate down to $M = 25$. To quantify these differences, we compute an uncertainty-weighted mean squared error (UW-MSE) between the generated and DNS-based $\zeta(4,\tau)$:
\begin{equation}
    \mathrm{UW\text{-}MSE} = \frac{\int[\zeta(4,\tau)-\zeta^{(\mathrm{DNS})}(4,\tau)]^2/\sigma^2(\zeta^{(\mathrm{DNS})}(4,\tau))\,d\tau}{\int 1/\sigma^2(\zeta^{(\mathrm{DNS})}(4,\tau))\,d\tau},
\end{equation}
where $\sigma^2(\zeta^{(\mathrm{DNS})}(4,\tau))$ denotes the variance of the DNS local slope at each scale $\tau$, computed over 30 batches spanning all velocity components. Fig.~\ref{fig:step_reduction}(c) shows the resulting UW-MSE as a function of $M$. Both models maintain low UW-MSE from the full 800 steps down to $M = 100$, but as $M$ decreases further, DDPM exhibits increasing error, while DDIM maintains low error down to $M = 25$. At $M = 5$, both models exhibit a substantial breakdown in multiscale accuracy, as reflected by a sharp rise in UW-MSE.

This result highlights DDIM’s robustness under aggressive step reduction and its promise for efficient Lagrangian turbulence generation. The contrasting behaviors of DDIM and DDPM can be attributed to their treatment of stochasticity: DDPM injects random noise at each reverse step, which facilitates mode exploration during full-length generation but may lead to error accumulation when the number of steps is reduced. In contrast, DDIM uses a deterministic mapping from the initial noise to the output, avoiding intermediate randomness and yielding more stable generation under shorter schedules. This deterministic formulation likely contributes to DDIM’s superior performance in reduced-step regimes.

\section{Conclusions}\label{sec:conclusions}

Building on recent advances in diffusion-based generative modeling of Lagrangian turbulence, this study examines three key aspects of diffusion models: their robustness across network architectures, the latent signatures of extreme events under DDIM sampling, and the trade-off between sampling efficiency and statistical fidelity. We show that, under shared sampling randomness, U-Net and transformer-based diffusion models generate highly correlated Lagrangian trajectories, indicating strong architectural consistency at the trajectory level. When assessing statistical accuracy across an ensemble of trajectories, the U-Net model performs well at all temporal scales across both DDPM and DDIM sampling schemes, while the transformer tends to underestimate small-scale intermittency---likely due to the absence of architectural tuning in this work.

To gain insight into the emergence of extreme events in diffusion-based generation, we analyzed the initial latent noise under DDIM sampling. Its deterministic mapping enables tracing output trajectories back to input noise. We found that large acceleration bursts consistently align with localized structures in the latent input, suggesting that rare events are encoded by specific variations in the generative prior.

Finally, we explored accelerated trajectory generation via reduced-step sampling schedules. Both DDPM and DDIM achieve substantial speedups under step reduction, but DDIM remains significantly more robust when the number of steps is aggressively reduced. With as few as 25 steps---compared to the original 800-step schedule---DDIM preserves multiscale statistical accuracy, whereas DDPM exhibits noticeable degradation at small scales. These results underscore the advantage of DDIM for efficient and scalable trajectory synthesis.

Together, these results highlight the potential of diffusion models as robust and interpretable tools for generating realistic Lagrangian turbulence. They also point toward several promising directions for future research, such as improving small-scale fidelity through architectural optimization, which is critical for representing intermittent dynamics and maintaining distributional richness. Other important directions include the controlled generation of rare events and scalable synthesis for larger datasets and higher-Reynolds-number turbulence.

\section*{Data and Code Availability}

The Lagrangian trajectory dataset used in this work, including both particle positions and velocities, is publicly available via the open-access Smart-TURB portal at \url{http://smart-turb.roma2.infn.it}~\citep{biferale2023turb}. The code for training the U-Net-based diffusion model and generating synthetic trajectories is available at \url{https://github.com/SmartTURB/diffusion-lagr}~\citep{SmartTURB2024DiffusionLagr}, and the code for the DiT-based diffusion model used in this study is available at \url{https://github.com/SmartTURB/transf-DM-lagr}.

% ----- ----- ----- ----- ----- ----- ----- ----- ----- -----  |

\section*{Acknowledgements}

We thank Antonio Celani and Mauro Sbragaglia for useful discussions. This work was supported by the European Research Council (ERC) under the European Union’s Horizon 2020 research and innovation programme Smart-TURB (Grant Agreement No. 882340). FT has received financial support from the CNRS through the MITI interdisciplinary programs through its exploratory research program.

% ----- ----- ----- ----- ----- ----- ----- ----- ----- -----  |

\appendix

\section{Derivation of Reverse Process Coefficients}
\label{app:app1}

This section derives the closed-form expressions for the reverse process coefficients $\omega_n$ and $\rho_n$ in Eq.~\eqref{eq:sol}. We start from the assumed Gaussian form of the reverse transition in Eq.~\eqref{eq:general_back}, and compute the marginal distribution $q(\mathcal{V}_{n-1}|\mathcal{V}_0)$ in two ways. First, from Eq.~\eqref{eq:marginal}, we know that
\begin{equation}
\label{eq:app_marg}
q(\mathcal{V}_{n-1}|\mathcal{V}_0) = \mathcal{N}(\sqrt{\bar{\alpha}_{n-1}} \mathcal{V}_0, (1 - \bar{\alpha}_{n-1})\bm{I}).
\end{equation}

Alternatively, using Eq.~\eqref{eq:general_back} and marginalizing over $\mathcal{V}_n$, we compute the same quantity as:
\begin{equation}
\label{eq:marginalization}
q(\mathcal{V}_{n-1}|\mathcal{V}_0) = \int q(\mathcal{V}_{n-1}|\mathcal{V}_n, \mathcal{V}_0) \, q(\mathcal{V}_n|\mathcal{V}_0)\, d\mathcal{V}_n.
\end{equation}
Using the Gaussian forms of both terms:
\begin{align*}
q(\mathcal{V}_{n-1}|\mathcal{V}_n, \mathcal{V}_0) &= \mathcal{N}(\omega_n \mathcal{V}_n + \rho_n \mathcal{V}_0, \sigma_n^2 \bm{I}), \\
q(\mathcal{V}_n|\mathcal{V}_0) &= \mathcal{N}(\sqrt{\bar{\alpha}_n} \mathcal{V}_0, (1 - \bar{\alpha}_n) \bm{I}),
\end{align*}
the integrand of Eq.~\eqref{eq:marginalization} becomes the product of two Gaussians, which can be written as:
\begin{align*}
&\exp\left\{ -\frac{1}{2\sigma_n^2} \left\| \mathcal{V}_{n-1} - (\omega_n \mathcal{V}_n + \rho_n \mathcal{V}_0) \right\|^2 \right\} \\
&\quad \times \exp\left\{ -\frac{1}{2(1 - \bar{\alpha}_n)} \left\| \mathcal{V}_n - \sqrt{\bar{\alpha}_n} \mathcal{V}_0 \right\|^2 \right\}.
\end{align*}
Combining the exponents and completing the square in $\mathcal{V}_n$ yields a quadratic form:
\begin{align*}
-\frac{1}{2(1 - \bar{\alpha}_n)} \Bigg[
&\left(1 + \frac{1 - \bar{\alpha}_n}{\sigma_n^2} \omega_n^2 \right) \|\mathcal{V}_n\|^2
- 2 \left( \sqrt{\bar{\alpha}_n} \mathcal{V}_0 + \frac{1 - \bar{\alpha}_n}{\sigma_n^2} \omega_n (\mathcal{V}_{n-1} - \rho_n \mathcal{V}_0) \right) \cdot \mathcal{V}_n \\
&+ \bar{\alpha}_n \|\mathcal{V}_0\|^2 + \frac{1 - \bar{\alpha}_n}{\sigma_n^2} \|\mathcal{V}_{n-1} - \rho_n \mathcal{V}_0\|^2 \Bigg].
\end{align*}
Letting $\lambda_n := 1 + \frac{1 - \bar{\alpha}_n}{\sigma_n^2} \omega_n^2$, integrating out $\mathcal{V}_n$ results in:
\begin{align*}
q(\mathcal{V}_{n-1}|\mathcal{V}_0) \propto \exp\Bigg\{ -\frac{1}{2(1 - \bar{\alpha}_n)} \Big[
&\bar{\alpha}_n \|\mathcal{V}_0\|^2 + \frac{1 - \bar{\alpha}_n}{\sigma_n^2} \|\mathcal{V}_{n-1} - \rho_n \mathcal{V}_0\|^2 \\
&- \frac{1}{\lambda_n} \left\| \sqrt{\bar{\alpha}_n} \mathcal{V}_0 + \frac{1 - \bar{\alpha}_n}{\sigma_n^2} \omega_n (\mathcal{V}_{n-1} - \rho_n \mathcal{V}_0) \right\|^2 \Big] \Bigg\}.
\end{align*}

We isolate all terms involving $\mathcal{V}_{n-1}$ and match this expression to the target form in Eq.~\eqref{eq:app_marg}, which yields the following system:
\begin{equation}
\left\{
\begin{aligned}
\frac{1 - \bar{\alpha}_{n-1}}{\sigma_n^2} \left( 1 - \frac{1 - \bar{\alpha}_n}{\lambda_n \sigma_n^2} \omega_n^2 \right) &= 1, \\
\sqrt{\bar{\alpha}_n} \cdot \frac{1 - \bar{\alpha}_{n-1}}{\sigma_n^2} \cdot \frac{\omega_n}{\lambda_n} + \rho_n &= \sqrt{\bar{\alpha}_{n-1}}.
\end{aligned}
\right.
\end{equation}
Solving this system for $\omega_n > 0$ yields the closed-form expressions:
\begin{align*}
\omega_n &= \sqrt{\frac{1 - \bar{\alpha}_{n-1} - \sigma_n^2}{1 - \bar{\alpha}_n}}, \\
\rho_n &= \sqrt{\bar{\alpha}_{n-1}} - \sqrt{\bar{\alpha}_n} \, \omega_n.
\end{align*}

% ----- ----- ----- ----- ----- ----- ----- ----- ----- -----  |

\bibliographystyle{elsarticle-harv} 
\bibliography{main}

\begin{thebibliography}{34}
\expandafter\ifx\csname natexlab\endcsname\relax\def\natexlab#1{#1}\fi
\providecommand{\url}[1]{\texttt{#1}}
\providecommand{\href}[2]{#2}
\providecommand{\path}[1]{#1}
\providecommand{\DOIprefix}{doi:}
\providecommand{\ArXivprefix}{arXiv:}
\providecommand{\URLprefix}{URL: }
\providecommand{\Pubmedprefix}{pmid:}
\providecommand{\doi}[1]{\href{http://dx.doi.org/#1}{\path{#1}}}
\providecommand{\Pubmed}[1]{\href{pmid:#1}{\path{#1}}}
\providecommand{\bibinfo}[2]{#2}
\ifx\xfnm\relax \def\xfnm[#1]{\unskip,\space#1}\fi
%Type = Article
\bibitem[{Arneodo et~al.(1998)Arneodo, Bacry and Muzy}]{arneodo1998random}
\bibinfo{author}{Arneodo, A.}, \bibinfo{author}{Bacry, E.},
  \bibinfo{author}{Muzy, J.F.}, \bibinfo{year}{1998}.
\newblock \bibinfo{title}{Random cascades on wavelet dyadic trees}.
\newblock \bibinfo{journal}{Journal of Mathematical Physics}
  \bibinfo{volume}{39}, \bibinfo{pages}{4142--4164}.
%Type = Article
\bibitem[{Arn{\'e}odo et~al.(2008)Arn{\'e}odo, Benzi, Berg, Biferale,
  Bodenschatz, Busse, Calzavarini, Castaing, Cencini, Chevillard
  et~al.}]{arneodo2008universal}
\bibinfo{author}{Arn{\'e}odo, A.}, \bibinfo{author}{Benzi, R.},
  \bibinfo{author}{Berg, J.}, \bibinfo{author}{Biferale, L.},
  \bibinfo{author}{Bodenschatz, E.}, \bibinfo{author}{Busse, A.},
  \bibinfo{author}{Calzavarini, E.}, \bibinfo{author}{Castaing, B.},
  \bibinfo{author}{Cencini, M.}, \bibinfo{author}{Chevillard, L.}, et~al.,
  \bibinfo{year}{2008}.
\newblock \bibinfo{title}{Universal intermittent properties of particle
  trajectories in highly turbulent flows}.
\newblock \bibinfo{journal}{Physical review letters} \bibinfo{volume}{100},
  \bibinfo{pages}{254504}.
%Type = Article
\bibitem[{Bacry and Muzy(2003)}]{bacry2003log}
\bibinfo{author}{Bacry, E.}, \bibinfo{author}{Muzy, J.F.},
  \bibinfo{year}{2003}.
\newblock \bibinfo{title}{Log-infinitely divisible multifractal processes}.
\newblock \bibinfo{journal}{Communications in Mathematical Physics}
  \bibinfo{volume}{236}, \bibinfo{pages}{449--475}.
%Type = Article
\bibitem[{Benzi et~al.(1993)Benzi, Ciliberto, Tripiccione, Baudet, Massaioli
  and Succi}]{benzi1993extended}
\bibinfo{author}{Benzi, R.}, \bibinfo{author}{Ciliberto, S.},
  \bibinfo{author}{Tripiccione, R.}, \bibinfo{author}{Baudet, C.},
  \bibinfo{author}{Massaioli, F.}, \bibinfo{author}{Succi, S.},
  \bibinfo{year}{1993}.
\newblock \bibinfo{title}{Extended self-similarity in turbulent flows}.
\newblock \bibinfo{journal}{Physical review E} \bibinfo{volume}{48},
  \bibinfo{pages}{R29}.
%Type = Article
\bibitem[{Biferale et~al.(1998)Biferale, Boffetta, Celani, Crisanti and
  Vulpiani}]{biferale1998mimicking}
\bibinfo{author}{Biferale, L.}, \bibinfo{author}{Boffetta, G.},
  \bibinfo{author}{Celani, A.}, \bibinfo{author}{Crisanti, A.},
  \bibinfo{author}{Vulpiani, A.}, \bibinfo{year}{1998}.
\newblock \bibinfo{title}{Mimicking a turbulent signal: Sequential multiaffine
  processes}.
\newblock \bibinfo{journal}{Physical Review E} \bibinfo{volume}{57},
  \bibinfo{pages}{R6261}.
%Type = Article
\bibitem[{Biferale et~al.(2004)Biferale, Boffetta, Celani, Devenish, Lanotte
  and Toschi}]{biferale2004multifractal}
\bibinfo{author}{Biferale, L.}, \bibinfo{author}{Boffetta, G.},
  \bibinfo{author}{Celani, A.}, \bibinfo{author}{Devenish, B.},
  \bibinfo{author}{Lanotte, A.}, \bibinfo{author}{Toschi, F.},
  \bibinfo{year}{2004}.
\newblock \bibinfo{title}{Multifractal statistics of lagrangian velocity and
  acceleration in turbulence}.
\newblock \bibinfo{journal}{Physical review letters} \bibinfo{volume}{93},
  \bibinfo{pages}{064502}.
%Type = Article
\bibitem[{Biferale et~al.(2023)Biferale, Bonaccorso, Buzzicotti and
  Calascibetta}]{biferale2023turb}
\bibinfo{author}{Biferale, L.}, \bibinfo{author}{Bonaccorso, F.},
  \bibinfo{author}{Buzzicotti, M.}, \bibinfo{author}{Calascibetta, C.},
  \bibinfo{year}{2023}.
\newblock \bibinfo{title}{Turb-lagr. a database of 3d lagrangian trajectories
  in homogeneous and isotropic turbulence}.
\newblock \bibinfo{journal}{arXiv preprint arXiv:2303.08662} .
%Type = Article
\bibitem[{Buzzicotti(2023)}]{buzzicotti2023data}
\bibinfo{author}{Buzzicotti, M.}, \bibinfo{year}{2023}.
\newblock \bibinfo{title}{Data reconstruction for complex flows using ai:
  Recent progress, obstacles, and perspectives}.
\newblock \bibinfo{journal}{Europhysics Letters} \bibinfo{volume}{142},
  \bibinfo{pages}{23001}.
%Type = Article
\bibitem[{Calascibetta et~al.(2023)Calascibetta, Biferale, Borra, Celani and
  Cencini}]{calascibetta2023optimal}
\bibinfo{author}{Calascibetta, C.}, \bibinfo{author}{Biferale, L.},
  \bibinfo{author}{Borra, F.}, \bibinfo{author}{Celani, A.},
  \bibinfo{author}{Cencini, M.}, \bibinfo{year}{2023}.
\newblock \bibinfo{title}{Optimal tracking strategies in a turbulent flow}.
\newblock \bibinfo{journal}{Communications Physics} \bibinfo{volume}{6},
  \bibinfo{pages}{256}.
%Type = Article
\bibitem[{Dhariwal and Nichol(2021)}]{dhariwal2021diffusion}
\bibinfo{author}{Dhariwal, P.}, \bibinfo{author}{Nichol, A.},
  \bibinfo{year}{2021}.
\newblock \bibinfo{title}{Diffusion models beat gans on image synthesis}.
\newblock \bibinfo{journal}{Advances in neural information processing systems}
  \bibinfo{volume}{34}, \bibinfo{pages}{8780--8794}.
%Type = Article
\bibitem[{Dosovitskiy et~al.(2020)Dosovitskiy, Beyer, Kolesnikov, Weissenborn,
  Zhai, Unterthiner, Dehghani, Minderer, Heigold, Gelly
  et~al.}]{dosovitskiy2020image}
\bibinfo{author}{Dosovitskiy, A.}, \bibinfo{author}{Beyer, L.},
  \bibinfo{author}{Kolesnikov, A.}, \bibinfo{author}{Weissenborn, D.},
  \bibinfo{author}{Zhai, X.}, \bibinfo{author}{Unterthiner, T.},
  \bibinfo{author}{Dehghani, M.}, \bibinfo{author}{Minderer, M.},
  \bibinfo{author}{Heigold, G.}, \bibinfo{author}{Gelly, S.}, et~al.,
  \bibinfo{year}{2020}.
\newblock \bibinfo{title}{An image is worth 16x16 words: Transformers for image
  recognition at scale}.
\newblock \bibinfo{journal}{arXiv preprint arXiv:2010.11929} .
%Type = Article
\bibitem[{Ho et~al.(2020)Ho, Jain and Abbeel}]{ho2020denoising}
\bibinfo{author}{Ho, J.}, \bibinfo{author}{Jain, A.}, \bibinfo{author}{Abbeel,
  P.}, \bibinfo{year}{2020}.
\newblock \bibinfo{title}{Denoising diffusion probabilistic models}.
\newblock \bibinfo{journal}{Advances in neural information processing systems}
  \bibinfo{volume}{33}, \bibinfo{pages}{6840--6851}.
%Type = Article
\bibitem[{Kadkhodaie et~al.(2023)Kadkhodaie, Guth, Simoncelli and
  Mallat}]{kadkhodaie2023generalization}
\bibinfo{author}{Kadkhodaie, Z.}, \bibinfo{author}{Guth, F.},
  \bibinfo{author}{Simoncelli, E.P.}, \bibinfo{author}{Mallat, S.},
  \bibinfo{year}{2023}.
\newblock \bibinfo{title}{Generalization in diffusion models arises from
  geometry-adaptive harmonic representations}.
\newblock \bibinfo{journal}{arXiv preprint arXiv:2310.02557} .
%Type = Article
\bibitem[{La~Porta et~al.(2001)La~Porta, Voth, Crawford, Alexander and
  Bodenschatz}]{la2001fluid}
\bibinfo{author}{La~Porta, A.}, \bibinfo{author}{Voth, G.A.},
  \bibinfo{author}{Crawford, A.M.}, \bibinfo{author}{Alexander, J.},
  \bibinfo{author}{Bodenschatz, E.}, \bibinfo{year}{2001}.
\newblock \bibinfo{title}{Fluid particle accelerations in fully developed
  turbulence}.
\newblock \bibinfo{journal}{Nature} \bibinfo{volume}{409},
  \bibinfo{pages}{1017--1019}.
%Type = Article
\bibitem[{Li et~al.(2024a)Li, Biferale, Bonaccorso, Buzzicotti and
  Centurioni}]{li2024stochastic}
\bibinfo{author}{Li, T.}, \bibinfo{author}{Biferale, L.},
  \bibinfo{author}{Bonaccorso, F.}, \bibinfo{author}{Buzzicotti, M.},
  \bibinfo{author}{Centurioni, L.}, \bibinfo{year}{2024}a.
\newblock \bibinfo{title}{Stochastic reconstruction of gappy lagrangian
  turbulent signals by conditional diffusion models}.
\newblock \bibinfo{journal}{arXiv preprint arXiv:2410.23971} .
%Type = Misc
\bibitem[{Li et~al.(2024b)Li, Biferale, Bonaccorso, Scarpolini and
  Buzzicotti}]{SmartTURB2024DiffusionLagr}
\bibinfo{author}{Li, T.}, \bibinfo{author}{Biferale, L.},
  \bibinfo{author}{Bonaccorso, F.}, \bibinfo{author}{Scarpolini, M.A.},
  \bibinfo{author}{Buzzicotti, M.}, \bibinfo{year}{2024}b.
\newblock \bibinfo{title}{Smartturb/diffusion-lagr: stable}.
\newblock \URLprefix \url{https://doi.org/10.5281/zenodo.10563386},
  \DOIprefix\doi{10.5281/zenodo.10563386}.
%Type = Article
\bibitem[{Li et~al.(2024c)Li, Biferale, Bonaccorso, Scarpolini and
  Buzzicotti}]{li2024synthetic}
\bibinfo{author}{Li, T.}, \bibinfo{author}{Biferale, L.},
  \bibinfo{author}{Bonaccorso, F.}, \bibinfo{author}{Scarpolini, M.A.},
  \bibinfo{author}{Buzzicotti, M.}, \bibinfo{year}{2024}c.
\newblock \bibinfo{title}{Synthetic lagrangian turbulence by generative
  diffusion models}.
\newblock \bibinfo{journal}{Nature Machine Intelligence} \bibinfo{volume}{6},
  \bibinfo{pages}{393--403}.
%Type = Article
\bibitem[{Li et~al.(2023)Li, Lanotte, Buzzicotti, Bonaccorso and
  Biferale}]{li2023multi}
\bibinfo{author}{Li, T.}, \bibinfo{author}{Lanotte, A.S.},
  \bibinfo{author}{Buzzicotti, M.}, \bibinfo{author}{Bonaccorso, F.},
  \bibinfo{author}{Biferale, L.}, \bibinfo{year}{2023}.
\newblock \bibinfo{title}{Multi-scale reconstruction of turbulent rotating
  flows with generative diffusion models}.
\newblock \bibinfo{journal}{Atmosphere} \bibinfo{volume}{15},
  \bibinfo{pages}{60}.
%Type = Article
\bibitem[{Li et~al.(2024d)Li, Tommasi, Buzzicotti, Bonaccorso and
  Biferale}]{li2024generative}
\bibinfo{author}{Li, T.}, \bibinfo{author}{Tommasi, S.},
  \bibinfo{author}{Buzzicotti, M.}, \bibinfo{author}{Bonaccorso, F.},
  \bibinfo{author}{Biferale, L.}, \bibinfo{year}{2024}d.
\newblock \bibinfo{title}{Generative diffusion models for synthetic
  trajectories of heavy and light particles in turbulence}.
\newblock \bibinfo{journal}{International Journal of Multiphase Flow}
  \bibinfo{volume}{181}, \bibinfo{pages}{104980}.
%Type = Article
\bibitem[{Loshchilov and Hutter(2017)}]{loshchilov2017decoupled}
\bibinfo{author}{Loshchilov, I.}, \bibinfo{author}{Hutter, F.},
  \bibinfo{year}{2017}.
\newblock \bibinfo{title}{Decoupled weight decay regularization}.
\newblock \bibinfo{journal}{arXiv preprint arXiv:1711.05101} .
%Type = Article
\bibitem[{L{\"u}bke et~al.(2023)L{\"u}bke, Friedrich and
  Grauer}]{lubke2023stochastic}
\bibinfo{author}{L{\"u}bke, J.}, \bibinfo{author}{Friedrich, J.},
  \bibinfo{author}{Grauer, R.}, \bibinfo{year}{2023}.
\newblock \bibinfo{title}{Stochastic interpolation of sparsely sampled time
  series by a superstatistical random process and its synthesis in fourier and
  wavelet space}.
\newblock \bibinfo{journal}{Journal of Physics: Complexity}
  \bibinfo{volume}{4}, \bibinfo{pages}{015005}.
%Type = Article
\bibitem[{Martin et~al.(2025)Martin, L{\"u}bke, Li, Buzzicotti, Grauer and
  Biferale}]{martin2025generation}
\bibinfo{author}{Martin, J.}, \bibinfo{author}{L{\"u}bke, J.},
  \bibinfo{author}{Li, T.}, \bibinfo{author}{Buzzicotti, M.},
  \bibinfo{author}{Grauer, R.}, \bibinfo{author}{Biferale, L.},
  \bibinfo{year}{2025}.
\newblock \bibinfo{title}{Generation of cosmic-ray trajectories by a diffusion
  model trained on test particles in 3d magnetohydrodynamic turbulence}.
\newblock \bibinfo{journal}{The Astrophysical Journal Supplement Series}
  \bibinfo{volume}{277}, \bibinfo{pages}{48}.
%Type = Article
\bibitem[{Mordant et~al.(2001)Mordant, Metz, Michel and
  Pinton}]{mordant2001measurement}
\bibinfo{author}{Mordant, N.}, \bibinfo{author}{Metz, P.},
  \bibinfo{author}{Michel, O.}, \bibinfo{author}{Pinton, J.F.},
  \bibinfo{year}{2001}.
\newblock \bibinfo{title}{Measurement of lagrangian velocity in fully developed
  turbulence}.
\newblock \bibinfo{journal}{Physical Review Letters} \bibinfo{volume}{87},
  \bibinfo{pages}{214501}.
%Type = Inproceedings
\bibitem[{Nichol and Dhariwal(2021)}]{nichol2021improved}
\bibinfo{author}{Nichol, A.Q.}, \bibinfo{author}{Dhariwal, P.},
  \bibinfo{year}{2021}.
\newblock \bibinfo{title}{Improved denoising diffusion probabilistic models},
  in: \bibinfo{booktitle}{International conference on machine learning},
  \bibinfo{organization}{PMLR}. pp. \bibinfo{pages}{8162--8171}.
%Type = Inproceedings
\bibitem[{Peebles and Xie(2023)}]{peebles2023scalable}
\bibinfo{author}{Peebles, W.}, \bibinfo{author}{Xie, S.}, \bibinfo{year}{2023}.
\newblock \bibinfo{title}{Scalable diffusion models with transformers}, in:
  \bibinfo{booktitle}{Proceedings of the IEEE/CVF international conference on
  computer vision}, pp. \bibinfo{pages}{4195--4205}.
%Type = Article
\bibitem[{Pope(2011)}]{pope2011simple}
\bibinfo{author}{Pope, S.B.}, \bibinfo{year}{2011}.
\newblock \bibinfo{title}{Simple models of turbulent flows}.
\newblock \bibinfo{journal}{Physics of Fluids} \bibinfo{volume}{23}.
%Type = Inproceedings
\bibitem[{Ronneberger et~al.(2015)Ronneberger, Fischer and
  Brox}]{ronneberger2015u}
\bibinfo{author}{Ronneberger, O.}, \bibinfo{author}{Fischer, P.},
  \bibinfo{author}{Brox, T.}, \bibinfo{year}{2015}.
\newblock \bibinfo{title}{U-net: Convolutional networks for biomedical image
  segmentation}, in: \bibinfo{booktitle}{Medical image computing and
  computer-assisted intervention--MICCAI 2015: 18th international conference,
  Munich, Germany, October 5-9, 2015, proceedings, part III 18},
  \bibinfo{organization}{Springer}. pp. \bibinfo{pages}{234--241}.
%Type = Article
\bibitem[{Sawford(1991)}]{sawford1991reynolds}
\bibinfo{author}{Sawford, B.}, \bibinfo{year}{1991}.
\newblock \bibinfo{title}{Reynolds number effects in lagrangian stochastic
  models of turbulent dispersion}.
\newblock \bibinfo{journal}{Physics of Fluids A: Fluid Dynamics}
  \bibinfo{volume}{3}, \bibinfo{pages}{1577--1586}.
%Type = Article
\bibitem[{Sawford(2001)}]{sawford2001turbulent}
\bibinfo{author}{Sawford, B.}, \bibinfo{year}{2001}.
\newblock \bibinfo{title}{Turbulent relative dispersion}.
\newblock \bibinfo{journal}{Annual review of fluid mechanics}
  \bibinfo{volume}{33}, \bibinfo{pages}{289--317}.
%Type = Inproceedings
\bibitem[{Sohl-Dickstein et~al.(2015)Sohl-Dickstein, Weiss, Maheswaranathan and
  Ganguli}]{sohl2015deep}
\bibinfo{author}{Sohl-Dickstein, J.}, \bibinfo{author}{Weiss, E.},
  \bibinfo{author}{Maheswaranathan, N.}, \bibinfo{author}{Ganguli, S.},
  \bibinfo{year}{2015}.
\newblock \bibinfo{title}{Deep unsupervised learning using nonequilibrium
  thermodynamics}, in: \bibinfo{booktitle}{International conference on machine
  learning}, \bibinfo{organization}{pmlr}. pp. \bibinfo{pages}{2256--2265}.
%Type = Article
\bibitem[{Song et~al.(2020)Song, Meng and Ermon}]{song2020denoising}
\bibinfo{author}{Song, J.}, \bibinfo{author}{Meng, C.}, \bibinfo{author}{Ermon,
  S.}, \bibinfo{year}{2020}.
\newblock \bibinfo{title}{Denoising diffusion implicit models}.
\newblock \bibinfo{journal}{arXiv preprint arXiv:2010.02502} .
%Type = Article
\bibitem[{Toschi and Bodenschatz(2009)}]{toschi2009lagrangian}
\bibinfo{author}{Toschi, F.}, \bibinfo{author}{Bodenschatz, E.},
  \bibinfo{year}{2009}.
\newblock \bibinfo{title}{Lagrangian properties of particles in turbulence}.
\newblock \bibinfo{journal}{Annual review of fluid mechanics}
  \bibinfo{volume}{41}, \bibinfo{pages}{375--404}.
%Type = Article
\bibitem[{Viggiano et~al.(2020)Viggiano, Friedrich, Volk, Bourgoin, Cal and
  Chevillard}]{viggiano2020modelling}
\bibinfo{author}{Viggiano, B.}, \bibinfo{author}{Friedrich, J.},
  \bibinfo{author}{Volk, R.}, \bibinfo{author}{Bourgoin, M.},
  \bibinfo{author}{Cal, R.B.}, \bibinfo{author}{Chevillard, L.},
  \bibinfo{year}{2020}.
\newblock \bibinfo{title}{Modelling lagrangian velocity and acceleration in
  turbulent flows as infinitely differentiable stochastic processes}.
\newblock \bibinfo{journal}{Journal of Fluid Mechanics} \bibinfo{volume}{900},
  \bibinfo{pages}{A27}.
%Type = Article
\bibitem[{Yeung(2002)}]{yeung2002lagrangian}
\bibinfo{author}{Yeung, P.}, \bibinfo{year}{2002}.
\newblock \bibinfo{title}{Lagrangian investigations of turbulence}.
\newblock \bibinfo{journal}{Annual review of fluid mechanics}
  \bibinfo{volume}{34}, \bibinfo{pages}{115--142}.

\end{thebibliography}

\end{document}